\documentclass[a4paper,12pt]{scrartcl}
\usepackage[english]{babel}
\usepackage{xspace}
\usepackage[format=plain,labelfont={bf},font=small,width=0.9\textwidth]{caption}
\setkomafont{disposition}{\normalcolor\bfseries}
\usepackage{amsmath}
\usepackage{commath}
\usepackage[range-phrase={ - },range-units=single,multi-part-units=single,detect-all]{siunitx}
\usepackage[version=4]{mhchem}
\usepackage{graphicx}
\usepackage{csquotes}
\usepackage[
backend=biber,
hyperref=true,
style=numeric-comp,
sorting=none,
doi=false,
url=false,
eprint=false,
isbn=false,
maxnames=20,
giveninits=true,
date=year
]{biblatex}
\DefineBibliographyExtras{english}{} 
\DeclareFieldFormat[article]{volume}{\mkbibbold{#1}}
\DeclareFieldFormat{pages}{#1}
\renewbibmacro{in:}{}
\renewbibmacro*{volume+number+eid}{
  \printfield{volume}
  \setunit*{\addcomma\space}}
\usepackage{hyperref}
\hypersetup{
pdfproducer	={},
pdfcreator ={}
}
\usepackage{authblk}

\newcommand*{\eg}{\textit{e.g.}\@\xspace}
\newcommand*{\ie}{\textit{i.e.}\@\xspace}
\newcommand*{\etal}{\textit{et al.}\@\xspace}
\newcommand*{\cf}{\textit{cf.}\@\xspace}

\newcommand*{\via}{\textit{via }}

\begin{document}
\title{Cyano-functionalized Ag-bis-acetylide wires on Ag(110)}
\author[1]{Raphael Hellwig}
\author[1]{Martin Uphoff}
\author[1]{Yiqi Zhang}
\author[1]{Mateusz Paszkiewicz}
\author[1]{Liding Zhang}
\author[2]{Ping Du}
\author[2,3]{Mario Ruben}
\author[1,4]{Florian Klappenberger}
\author[1,5]{Johannes V. Barth}
\affil[1]{Physics Department E20, Technical University of Munich, D-85748 Garching, Germany}
\affil[2]{Institut für Nanotechnologie (INT), Karlsruher Institut für Technologie (KIT), D-76344 Eggenstein-Leopoldshafen, Germany}
\affil[3]{Institut de Physique et Chimie de Mat\'{e}riaux de Strasbourg (IPCMS), CNRS-Universit\'{e} de Strasbourg, F-67034 Strasbourg, France}
\affil[4]{florian.klappenberger@tum.de}
\affil[5]{jvb@tum.de}
\date{}
\maketitle

\begin{abstract}
Organometallic nanostructures are promising candidates for applications in optoelectronics, magnetism and catalysis. Our bottom-up approach employs a cyano-functionalized terminal alkyne species (CN-DETP) on the Ag$(110)$ surface to fabricate 2D domains of regularly stacked Ag-acetylide nanowires. We unravel their adsorption properties and give evidence to their organometallic character with the aid of complementary surface-sensitive techniques, \ie scanning tunneling microscopy, X-ray photoelectron spectroscopy and near-edge X-ray absorption fine-structure spectroscopy. Guided by the anisotropic $(110)$ surface, highly oriented nanowires form in two enantiomorphic domains of regularly stacked trans isomers, whereby the bifunctional design of CN-DETP gives rise to orthogonal bonding motifs. Based on STM imaging, we find high thermal stability of the Ag-bis-acetylide wires, without conversion into graphdiyne chains. Our approach based on orthogonal bifunctionalization and selective functional group recognition extends the toolbox of creating alkyne-based nanostructures at interfaces.
\end{abstract}

\newpage
\section{Introduction}
The bottom-up fabrication of low-dimensional nanostructures can be achieved by rational design of reactive molecular linkers and their deposition on single-crystalline metal surfaces. Depending on precursor characteristics and catalytic effects, different reaction pathways are employed, such as organometallic and covalent coupling, \eg Ullmann coupling \cite{Ruffieux2016,Klappenberger2018,Zhang2018,Moreno2018}. Organometallic complexes and metal-bis-acetylide motifs based on alkyne precursors allow the on-surface synthesis of complex structures including surface tesselations \cite{Hellwig2017,Zhang2018}. Furthermore, we have recently shown that lanthanide atoms catalyze alkyne deprotonation and covalent C-C linkage at temperatures down to \SI{130}{K} \cite{Hellwig2018}. While the sequence of alkyne deprotonation and linkage depends on the reaction type itself \cite{Bjoerk2016}, the intermediate steps of C-C bond formation frequently involve organometallic bonds with substrate atoms.  Based on the Ullmann-like coupling of halogen-functionalized precursors intermediate phases may interfere prior to covalent linkage, featuring organometallic bonds between dehalogenated endgroups and metal atoms \cite{Bieri2011,Cardenas2013,Giovannantonio2013}. Also for the linkage of poly-N-heterocyclic compounds, twofold metal-molecule coordination triggers intermediate phase formation prior to the creation of covalent chains \cite{Matena2008}.

Targeting the on-surface synthesis of extended molecular wires, the anisotropic (110) facet of noble metal substrates promotes a linear polymer growth mode different from the behavior of the (111) facet often leading to rotational domains. For instance, stable organometallic bonding between Cu atoms and saturated hydrocarbon moieties of porphyrin molecules was achieved during surface-assisted deprotonation on the Cu(110) surface \cite{Haq2011,Vasseur2016,Lin2017,Kalashnyk2017}. Furthermore, covalent catenation of an alkane precursor was accomplished by Zhong \etal on the reconstructed Au(110) surface, which constrains molecular orientation and diffusion, and bestows lower activation barriers compared to related isotropic metal facets \cite{Zhong2011}. While the  C-H bond scission of saturated carbon moieties requires elevated substrate temperatures to obtain organometallic (C-M-C) or covalent (C-C) carbon bonds, unsaturated alkyne functionalization appears to reduce activation barriers for covalent and organometallic coupling on silver substrates \cite{Zhang2012,Cirera2014,Liu2015b}.

Not only the choice of the substrate registry, but also the precursor design strongly influences the growth mode of molecular chains. Recently, we have shown that the precursor 4,4''-diethynyl-\lbrack 1,1':4',1''-terphenyl\rbrack -3,3''-dicarbonitrile (CN-DETP) containing secondary cyano sidegroups boosts the chemoselectivity toward linear molecular coupling motifs (butadiyne groups stemming from terminal alkyne homocoupling) on Ag(111), while the DETP monomer (without cyano (CN) sidegroups) results in the formation of various irregular coupling motifs involving orhtogonal interactions \cite{Klappenberger2018}.

Targeting further optimization of the linear polymer growth of nanochains, here we study bifunctionalized CN-DETP on an anisotropic surface, \ie the Ag(110) surface. We find that instead of the homocoupling reaction proceeding on Ag(111) \cite{Klappenberger2018}, the Ag(110) surface steers terminal alkyne coupling \via organometallic Ag-bis-acetliyde linkage. Our results are in line with other studies reporting on linear metal-bis-acetylide molecular wires based on terminal alkyne \cite{Liu2015b,Sun2016a} and terminal alkynyl \cite{Sun2016} coupling. Our study presents comprehensive X-ray spectroscopy (XS) results delineating adsorption, geometric, and organometallic bonding aspects in addition to STM.

When combining low-temperature scanning tunneling microscopy (STM) with syn-chrotron-based X-ray photoelectron spectroscopy (XPS) and near-edge X-ray absorption fine-structure (NEXAFS) measurements, our understanding on adsorption details is significantly extended by an element- and orbital-selective assessment of the underlying chemical, electronical and conformational aspects.

Furthermore, we employ a precursor with secondary functional groups mediating sidechain interactions pushing toward the formation of 2D domains. The latter are guided through supramolecular CN $\cdots$ H recognition between the conformer states of CN-DETP enantiomers. Based on the selective orthogonal interactions \cite{Hofmeier2005,Saha2013,Urgel2015} between terminal alkynes toward Ag adatoms and cyano groups toward comparatively weak electrostatic interactions with methine moieties, we identify a heterofunctionalized interface composed of regularly packed isochiral nanochains. Since the chains follow functional CN$\cdots$H recognition principles between interlinked enantiomers, the domains are comprised of isochiral trans isomers coordinated to a periodic array of Ag atoms.

Related organometallic structures take on great significance for a plethora of synthesis protocols in organic chemistry, targeting selective and energy-efficient synthesis strategies toward carbon-based structures. Silver-acetylide compounds, \ie a Ag atom binding to sp-hybridized carbon atoms, represent selective catalysts for the activation of versatile addition reactions taking place under gentle reaction conditions \cite{Halbes-Letinois2007,Yamamoto2008,Fang2015}. Furthermore, densely packed domains of metallopolymer chains represent useful interfacial systems for artificial light harvesting \cite{Wong2010} and for luminescent materials \cite{Yam2002}.

\section{Results and discussion}

\begin{figure}
	\begin{center}
		\includegraphics[width=0.9\textwidth]{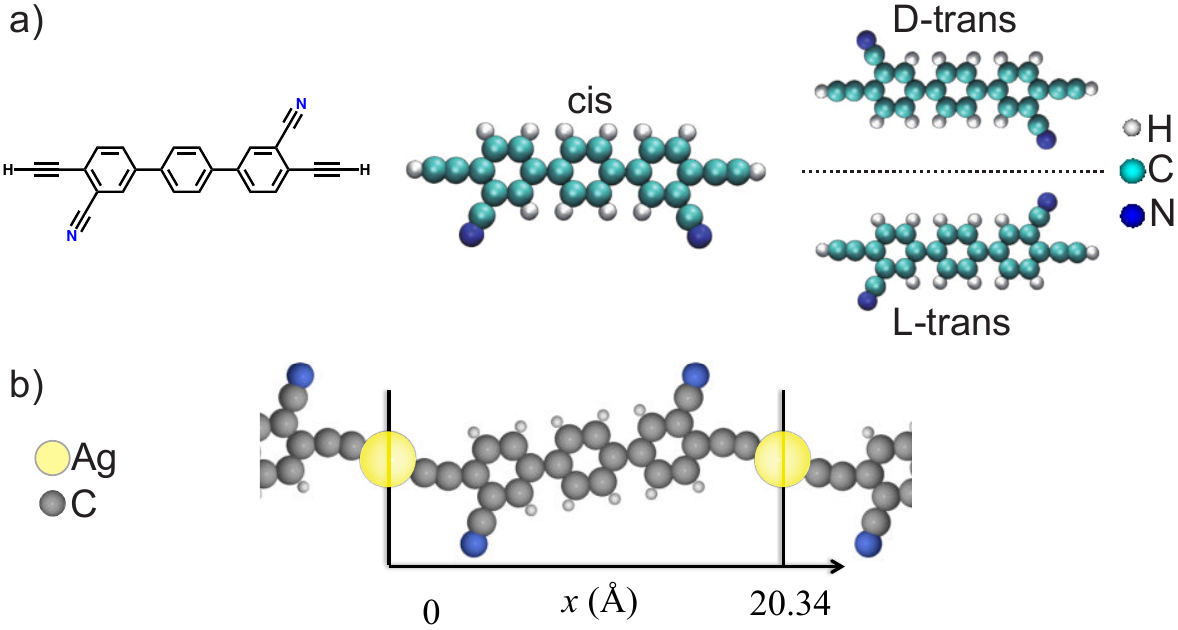}
	\end{center}
	\caption{Molecular model for CN-DETP. (a) Chemical structure model (left) and ball and stick model (right) of the pristine molecule. The labeling indicates the isomeric states of CN-DETP upon 2D confinement: cis, D-trans, L-trans.  (b) Gas-phase DFT model of organometallic Ag-bis-acetylide molecular chain \cite{jonas}. The cyano-phenylene groups can rotate around $\sigma$ bonds during molecular self-assembly.}
	\label{molecule}
\end{figure}

The chemical structure of CN-DETP is presented in the left part of Figure \ref{molecule}a: a \textit{para}-terphenyl backbone equipped with terminal alkyne endgroups and cyano sidegroups. Details on the chemical synthesis route are presented in ref. \cite{Klappenberger2018}. According to our previous results, the positioning of CN gropus entails 2D isomers \cite{Marschall2010,Marschall2010a}, while the terminal alkynes enable heat-induced on-surface coupling reactions \cite{Cirera2014}. The surface-adsorbed monomers are divided into a chiral D-trans and a L-trans enantiomers and an achiral cis species. These isomer species are represented by atomic ball and stick models (modeled \via Hyperchem software \cite{hyperchem}) in the right and middle part of Figure \ref{molecule}a, respectively. For sufficient thermal activation, the cis$\leftrightarrow$trans isomerization barrier is overcome by a \SI{180}{\degree} rotation of a CN-phenylene unit, thereby allowing chiral adaptation of molecular conformer states during molecular self-assembly. Higher thermal activation leads to the linkage of terminal alkynes, \ie homocoupling on Ag$(111)$ \cite{Klappenberger2018} and Ag-bis-acetylide linkage on the more reactive Ag(110) surface, as reported in the following. Figure \ref{molecule}b shows an organometallic nanochain simulated through gas-phase density functional theory (DFT) modeling \cite{jonas}. The incorporated L-trans isomers are interlinked through C-Ag-C bridges. 

\subsection{Organic phase}

\begin{figure}
	\begin{center}
		\includegraphics[width=0.9\textwidth]{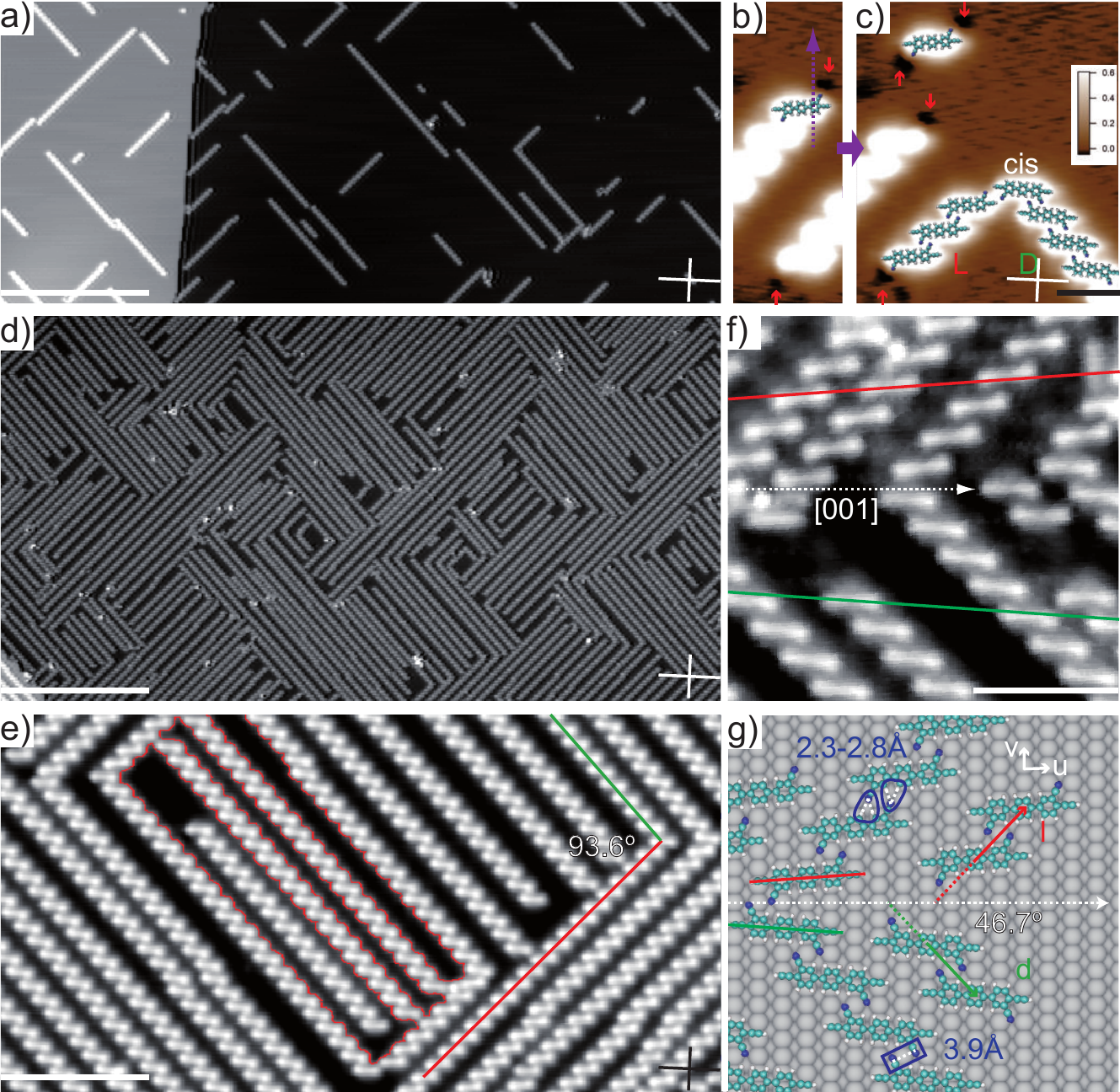}
	\end{center}
\caption{Supramolecular assemblies with low (a-c) and high (d-f) molecular coverage deposited at \SI{240}{K}. (a) STM image showing the chaining of monomers along two perpendicular propagation directions. (b,c) STM tip-guided isolation of one monomer, identified as L-trans unit according to CN-related depressions (red markers) and superimposed molecular models. STM contrast is enhanced and color-coded to emphasize the visibleness of depressions. (d,e) High-coverage preparation showing D- and L-trans ribbons enclosing rectangular voids. (f)  High-resolution STM image reveals molecular alignment.  (g) Adsorption model for the D-trans and L-trans ribbons. Monomer alignment (colored lines), surface unit cell vectors ($\vec u$ and $\vec v$) and molecular unit cell vectors ($\vec d$ and $\vec l$, each enclosing a \SI{46.7}{\degree} angle with $[001]$ direction (horizontal dashed line)) are indicated. C-N$\cdots$H and C-N$\cdots$N-C distances (numeric values) according to depicted dashed white lines within blue outlines. Tunneling parameters $V_t, I_t$: (a) \SI{1}{V}, \SI{0.1}{nA}; (b, c) ,\SI{0.01}{V}, \SI{1}{nA}; (d) \SI{0.1}{V}, \SI{0.1}{nA}; (e) \SI{-0.1}{V}, \SI{0.1}{nA}; (f) \SI{0.01}{V}, \SI{0.1}{nA}. Scale bars: (a) \SI{40}{nm}; (b, c) \SI{2}{nm}; (d) \SI{40}{nm}; (e) \SI{10}{nm}; (f) \SI{4}{nm}.}
	\label{organic}
\end{figure}

To explore the supramolecular assembly of CN-DETP monomers prior to heat-assisted alkyne reactions, we deposited a submonolayer (sub-ML) amount of monomers on the clean Ag(110) substrate kept at \SI{240}{K}. After cooling the sample down to STM imaging conditions at \SI{4.4}{K}, at which monomer diffusion and cyano-phenylene rotations are quenched, we observe the formation of one-dimensional (1D) ribbons running almost perpendicular with respect to each other. Figure \ref{organic}a shows well-defined ordering among on-surface formed ribbons comprised of dozens of molecules. Figure \ref{organic}c clearly shows the individual molecules imaged as rod-like protrusions. When enhancing the STM contrast (color-coding) in Figure \ref{organic}b, one notices depressions (red arrows) located next to the outer molecules terminating the 1D ribbons. To inspect an isolated monomer a terminal precursor was pulled away from the ribbon \via tip manipulation. For this, the STM tip was approached to the pristine Ag$(111)$ surface until reaching $I_t=\SI{55}{nA}$, followed by a displacement of the monomer under open-feedback conditions along the trajectory indicated by the purple dashed line. From the subsequent STM image (Figure \ref{organic}c), we infer that the depressions coincide with unbound cyano moieties. According to the superimposed molecular model, a L-trans monomer is identified. A similar depression-like feature in the STM contrast was reported by Vitali \etal for a molecule with deprotonated carboxylate and pyridin functional groups confined on Cu$(111)$. It was explained by modulations in the tunneling barrier due to a surface-induced dipole distribution in the vincinity of the electronegative moieties \cite{Vitali2008}. In our context, chemical sensitivity can be attributed to the STM tip resolving unbound cyano moieties, and hence identify isomer states (see ref. \cite{Klappenberger2018}). Accordingly, the absence of depression-like features next to interacting CN groups signals involvement in intermolecular interactions. From the location of the depressions found at the ribbons' terminations, we conclude that all its conformers express the same isomer state. Accordingly, all ribbons with positive (negative) inclination with respect to the $[001]$ direction (Figure \ref{organic}c, horizontal line of white cross) are formed by L-trans (D-trans) enantiomers. Occasionally, the terminating molecule of one ribbon binds to another ribbon, which implies a cis monomer without unbound cyano groups (no depressions).

Figure \ref{organic}d shows the ribbon phase at a higher molecular coverage giving rise to a maze-like nanostructure with enclosed voids. The STM image in Figure \ref{organic}e highlights that each corner of the rectangle contours (red outline) occurs at the intersection of a L- and a D-trans ribbon. Two opposing corners enclose an angle of \SI{93.6}{\degree}. From the high-resolution STM topograph in Figure \ref{organic}f, we conclude that D-trans (L-trans) molecules are slightly tilted ($\pm\SI{4}{\degree}$) with respect to the $[001]$ direction, depicted as dashed arrow representing the mirror axis.

Previously, Marschall \etal communicated the supramolecular assembly of the related linker $[1,1';4',1'']$-terphenyl-$3,3''$-dicarbonitrile that does not contain terminal alkyne moieties \cite{Marschall2010}. For sub-ML coverages on Ag$(111)$, a similar ribbon phase was observed, but with six molecular orientations (three pairs of D- and L-trans) according to the conformer's mirror symmetry with respect to the threefold dense-packed crystallographic directions of Ag$(111)$. Abbasi \etal rationalized the growth of extended D-trans (L-trans) ribbons on Ag(111) by a favorable mutual hydrogen bonding scenario combined with a particular isomerization mechanism \cite{Abbasi-Perez2014, Abbasi-Perez2015}. In agreement to our observations, a regular stacking of the molecules is reported, whereby the cis$\leftrightarrow$trans isomerization barrier for CN-phenyl rotations is overcome in order to optimize molecular recognition between trans enantiomers. Directional attractions between the cyano endgroups and neighboring organic ring moieties of adjacent enantiomers are quantified in the framework of the so-called PARI-interaction \cite{Arras2012}. Based on the STM findings, we suggest a commensurate adsorption model (Figure \ref{organic}), which will be corroborated through further STM data Figure \ref{organometallic}b, as explained below. As indicated by the depicted molecular models, the ribbons propagate along the vectors $\vec d$ and $\vec l$, which span the unit cell of the 2D self-assembly. In matrix notation, it reads
\begin{align*}
\left(\begin{array}{c} \vec{d} \\ \vec{l} \end{array}\right)=\left(\begin{array}{cc} 2 & -3 \\2 & 3\end{array}\right) \left(\begin{array}{c} \vec{u} \\ \vec{v} \end{array}\right)
\end{align*}
with respect to the unit cell vectors $\vec u$ and $\vec v$ of Ag$(110)$. The vectors are shown in Figure \ref{organic}g together with the C-N$\cdots$H and C-N$\cdots$N-C distances following from the model. The resulting monomer stacking along with the molecular alignment of D- and L-trans conformers (colored lines and \SI{46.7}{\degree} angle between horizontal dashed line and both propagation vectors, respectively; \cf Figure \ref{organic}g) agree with the enclosed angle of \SI{93}{\degree} between D-trans and L-trans ribbons in Figure \ref{organic}a,c and e.

\subsection{Organometallic phase}

\begin{figure}
	\begin{center}
		\includegraphics[width=0.9\textwidth]{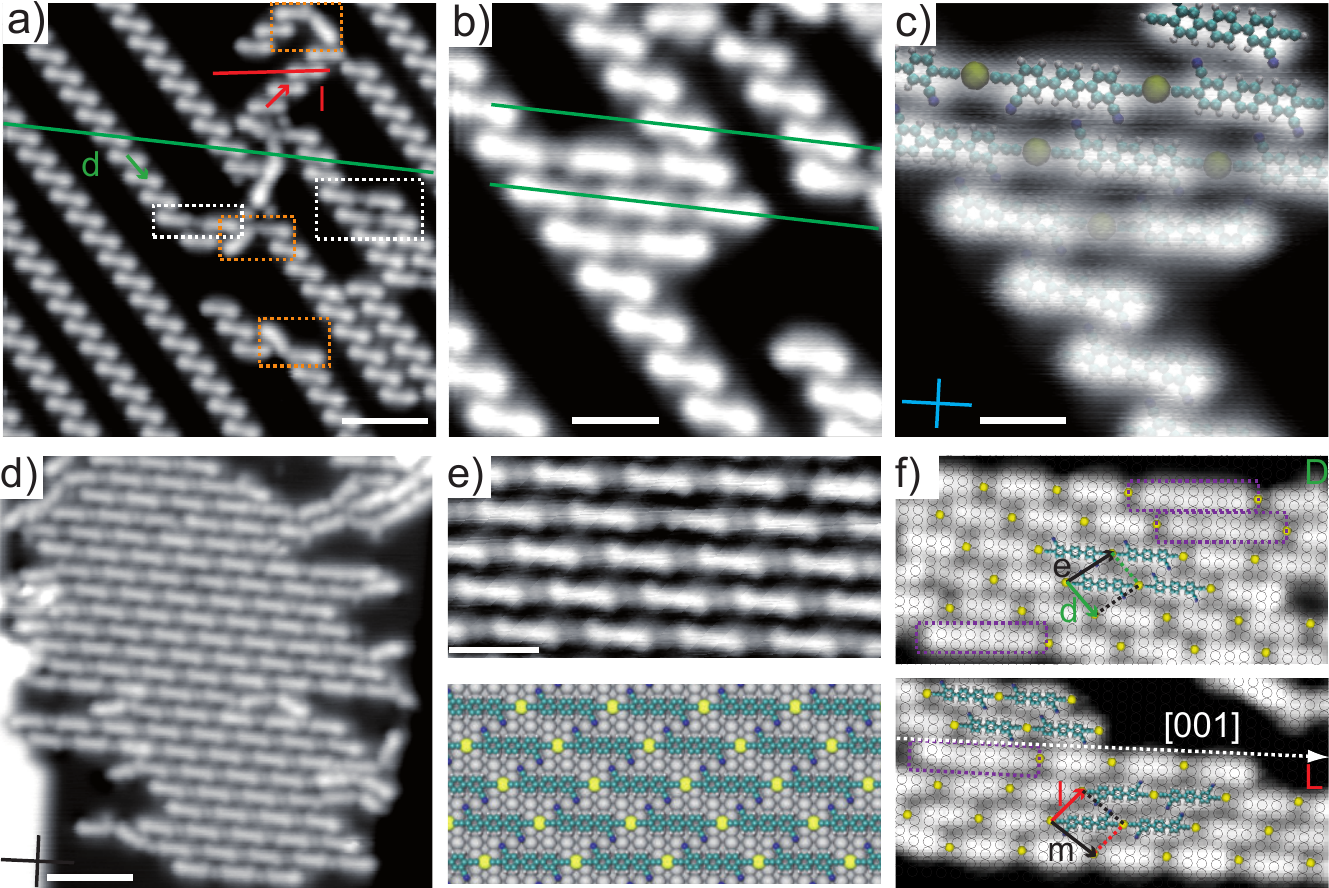}
	\end{center}
	\caption{Organometallic structures after sample annealing at \SI{300}{K} (a-c) and \SI{400}{K} (d-f). (a) D- and L-trans ribbons (according to molecular alignment (colored lines) and propagation vectors) coexisting with dimer structures (dashed outlines). (b) Organometallic dimer and trimers connecting two D-trans ribbons. (c) High-resolution STM image with superimposed models (molecules and Ag atoms). (d) Dense-packed stacking of 1D silver-acetylide wires within ordered islands. (e) High-resolution STM image of domain (top) and suggested model (bottom). (f) STM images of D- and L-trans domains with superimposed Ag(110) registry and coordinated Ag atoms (yellow). Superimposed molecular models display enantiopure D-trans (L-trans) chains and purple outlines depict covalent dimers (byproducts). Compare vectors of molecular unit cell with vectors in (a). Tunneling parameters $V_t, I_t$: (a) \SI{-0.1}{V}, \SI{0.1}{nA}; (b) \SI{0.1}{V}, \SI{0.1}{nA}; (c) \SI{0.01}{V}, \SI{0.1}{nA}; (d) \SI{0.05}{V}, \SI{0.3}{nA}; (e) \SI{0.005}{V}, \SI{0.5}{nA}; (f) \SI{0.01}{V}, \SI{0.1}{nA}. Scale bars: (a) \SI{4}{nm}; (b) \SI{2}{nm}; (c) \SI{1}{nm}; (d) \SI{4}{nm}; (e) \SI{2}{nm}.
}
	\label{organometallic}
\end{figure}

To explore the potential of surface-confined CN-DETP toward selective alkyne coupling reactions, we annealed the sample with the ribbon-like monomer phase at room temperature (\SI{300}{K}) and again carried out LT-STM measurements (Figure \ref{organometallic}). Besides the already known 1D ribbons, novel motifs can be observed in Figure \ref{organometallic}a, which are highlighted with dashed outlines. The STM image in Figure \ref{organometallic}b displays how monomers of adjacend ribbons become involved in head-to-tail bonding motifs with sphere-shaped protrusions along the molecular axis. From careful inspection of high-resolution STM data (Figure \ref{organometallic}c), the bonding distance appears larger than for covalent butadiyne bridges \cite{Cirera2014}. A scaled molecular model superimposed on top of the STM image strongly suggests the assumption of incorporated Ag atoms (yellow) binding to deprotonated alkynes (\cf Figure~\ref{molecule}b). For the C-Ag-C coupling motifs, we consider the tendency of Ag atoms to occupy the lower-lying fourfold hollow sites, in agreement with ref. \cite{Liu2015b} From the observed linkage of adjacent ribbons \via Ag-bis-acetylide linking (Figure~\ref{organometallic}b), we postulate that unreacted alkynes are located at bridge sites (next to hollow sites), as proposed in the adsorption model in Figure~\ref{organic}g.

A complete conversion from organic ribbons to silver-acetylide chains is accomplished through sample heating at temperatures between \num{350} and \SI{400}{K}. STM data (Figure \ref{organometallic}d) reveals dense-packed domains , which display molecular protrusions (bricks) and a periodic lattice of spherical spots. The uniform STM appearance of the interlinked molecules (Figure~\ref{organometallic}e) indicates that they express the same isomeric state (D-trans).
Accordingly, we propose an adsorption model (see bottom part of Figure~\ref{organometallic}e) with Ag atoms (yellow) centered at hollow sites of the Ag(110) registry, and deprotonated molecules forming an enantiopure D-trans domain. The coexistent L-trans domain is shown in the bottom part of Figure~\ref{organometallic}f. The model of the unit cells of both dense-packed domains separated by a white line along the $[$001$]$ direction (mirror axis) suggests that $\vec d$ and $\vec l$ are the same as within the organic phase (\cf Figure \ref{organometallic}a, Figure \ref{organic}g). From the superimposed Ag$(110)$ lattice model, one notices that all incorporated Ag atoms resite at fourfold hollow sites. They are assumed to be mobilized through evaporation from kinks and step edges during the annealing process \cite{Zambelli1998}. As a result of on-surface deprotonation, linear linking motifs reminiscent of $\sigma$ bonding are comprised of one Ag atom between two alkynyl groups.

The 2D superlattice of Ag atoms within the domains is rationalized by CN-mediated interchain recognition between stacked organometallic nanowires. We assume that electrostatic bonding between cyano moieties and aromatic hydrogen prevails for the dense-packed network. Since this kind of regular backbone-to-backbone bonding was already observed for the ribbon phase, the formerly introduced vectors $\vec d$ and $\vec l$ (\cf Figure \ref{organic}g) are retrieved within the unit cell of the two organometallic domains. 

The D- and L-trans domains obey mirror symmetry with respect to the $[001]$-direction (dashed arrow), \ie the line that connects embedded Ag centers (yellow) along an oligomeric chain, assuming that Ag centers coincide with hollow sites. This implies that the N atoms do not form metal-organic coordination bonds with mobilized Ag atoms, but favor supramolecular recognition between adjacent isomers, as observed for the purely molecular monomer organization. 

As apparent from Figure \ref{organometallic}c, the molecular orientations of pure CN-DETP and Ag-coordinated chains are very similar. We assume that the "orthogonality" of C-Ag-C and CN$\cdots$H interactions is required in order to establish an ordered 2D nanoarchitecture with organometallic wires stacked in a regular fashion. Following the previous matrix notation, the D-trans domain expresses the stacking of D-trans wires along the vector $\vec{d}$ of the pristine phase according to the matrix notation
\begin{align*}
\left(\begin{array}{c} \vec{d} \\ \vec{e} \end{array}\right)=\left(\begin{array}{cc} 2 & -3 \\3 & 3\end{array}\right) \left(\begin{array}{c} \vec{u} \\ \vec{v} \end{array}\right),
\end{align*} 
while the L-trans domain composed of L-trans enantiomers reads
\begin{equation*}
\left(\begin{array}{c} \vec{l} \\ \vec{m} \end{array}\right)=\left(\begin{array}{cc} 2 & 3 \\3 & -3\end{array}\right) \left(\begin{array}{c} \vec{u} \\ \vec{v} \end{array}\right).
\end{equation*}

While the alkyne endgroups prefer an exact alignment of C-Ag-C oligomers along the underlying $[001]$ direction \cite{Liu2015b}, cyano-mediated interactions may lead to a different orientation of interlinked D- and L-trans isomers analogous to the monomers of the organic phase expressing $\pm\SI{4}{\degree}$ twist with respect to $[001]$ direction). Accordingly, we assume that the deviation of the chains’ alignment from the $[001]$ axis (top panel of Figure \ref{organometallic}e) reflects a balance between CN-mediated interchain interactions and adsorption site-selectivity of C-Ag-C units. Based on the gas-phase DFT model visualization in Figure \ref{molecule}b, \cite{jonas} the distance between two incorporated Ag atoms along the metallopolymer wire amounts to $\approx\SI{20.34}{\angstrom}$.
This numeric value nicely agrees with the lattice distance between the respective hollow sites (\SI{20.4}{\angstrom} according to Figure \ref{organometallic}f) confining twofold coordinated Ag atoms on Ag(110).

The isochiral character of the dense-packed domains was demonstrated by STM tip-assisted manipulation experiments and subsequent identification of CN-related depressions (see suppporting information (SI) Figure S1a,b). There, the STM image after the tip-induced displacement of a chain termination indicates a high bending flexibility of the Ag-acetylide bond within organometallic molecular wires. Remarkable bending capabilities were also observed for isolated cyano-functionalized graphdiyne wires created through terminal alkyne homcoupling on Ag(111) \cite{Klappenberger2018}.
We assign the high flexibility and structural integrity of these 1D polymer structures to the mechanical strength and bending flexibility of sp-unsaturated carbon atoms within diyne units (\ce{C#C-C#C}) \cite{Szafert2003} and Ag-bis-acetylide units (\ce{C#C-Ag-C#C}).

Besides organometallic dimers and trimers closely aligned along the molecular axis of the monomers (white outlines in Figure \ref{organometallic}a), we observe the coexistence of a small quantity of dimers expressing either a shorter dimer length (purple outlines in Figure \ref{organometallic}f) or kinked geometry (orange outlines in Figure \ref{organometallic}a). As discussed in the SI (see SI Figure S2), these byproducts are assigned to covalent C-C bond formation. These dimeric motifs infrequently occur within densely packed organometallic domains (\cf SI Figure S2 b,e).

\subsection{X-ray spectroscopy}
After unraveling the adsorption behavior and structural properties, the chemical aspects of purely organic and organometallic phases were inspected by synchrotron-based XPS experiments. On the basis of the STM-based preparation protocols mentioned above, the clean Ag(110) is exposed to a sub-ML amount of molecules at a sample temperature of \SI{200}{K}, and characterized both before and after performing annealing at \SI{400}{K}. While the informative value of STM experiments is limited to microscopic regions of the molecular adlayer, XPS-based results present ensemble properties averaged over the macroscopic spot size of the X-ray beam. To reduce monomer mobility and radiation damage, all spectroscopy experiments were conducted at a sample temperature of \SI{200}{K}.

\begin{figure}
	\begin{center}
\includegraphics[width=0.5\textwidth]{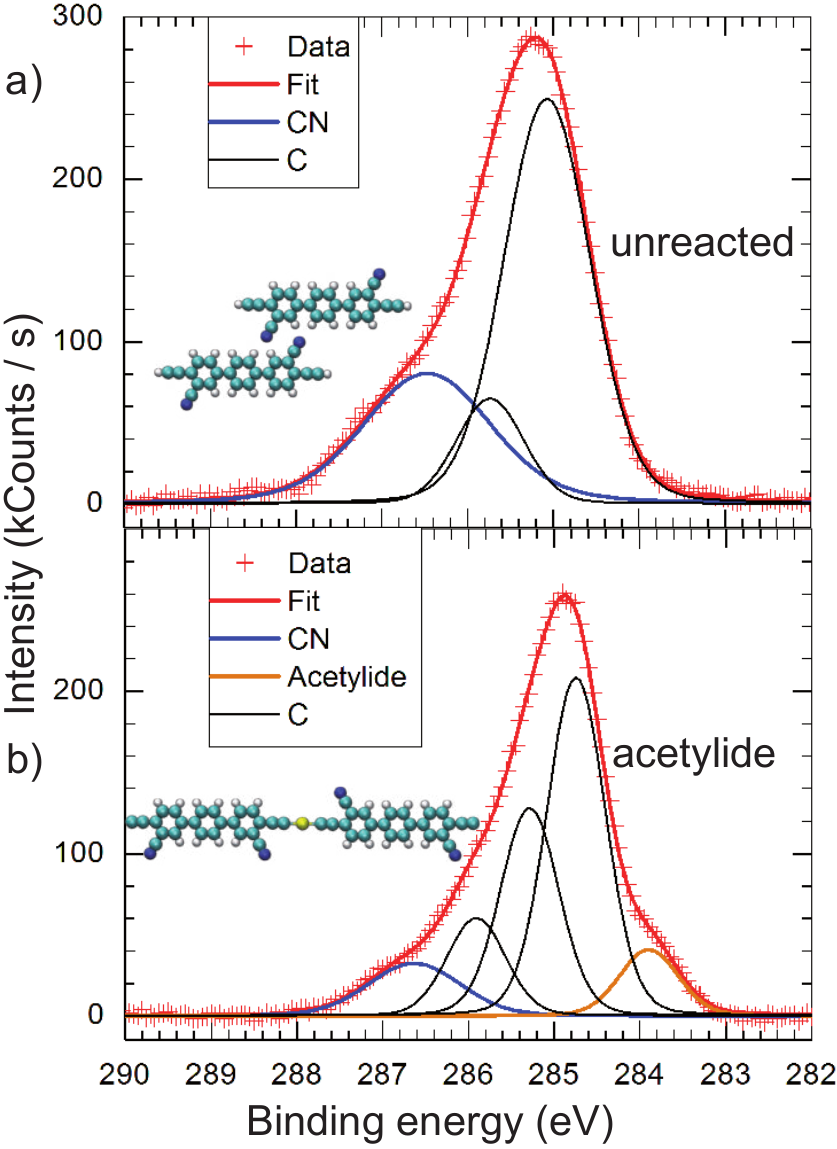}
	\end{center}
	\caption{C1s XPS characterization of pristine (a) and annealed sample (b) (pursuant to depicted molecular models). 
The comparison reveals a downshift in BE (\SIrange{285.2}{284.9}{eV}) and the emergence of a low-BE shoulder (\SI{283.9}{eV}) upon sample annealing at \SI{400}{K}.}
	\label{XPS}
\end{figure}

C1s XP spectra of the organic layer (\cf Figure \ref{organic}e) and annealed sample (\cf Figure \ref{organometallic}d) are presented in the top and bottom part of Figure \ref{XPS}. Since the experimental resolution is not sensitive enough to disentangle each chemical species contributing to the XPS profile (red data points) in Figure \ref{XPS}, the latter is approximated by a sum of individual Voigt fits (black curves) accounting for the different chemical carbon environments within \ce{C#N} and \ce{C#C}, as well as C(sp$^2$) within the monomer backbone. The high binding energy shoulder at \SI{286.5}{eV} (blue fitting curve) for the organic phase (top panel) represents the spectral contribution from the cyano endgroups. The maximum of the XPS profiles shift down from \SI{285.2}{eV} to \SI{284.9}{eV} after annealing the sample at \SI{400}{K} (bottom panel). For the latter, a low binding energy (BE) shoulder emerges at \SI{283.9}{eV} (orange fitting curve). Within the scope of a previous study on the alkyne derivative 1,3,5-triethynyl-benzene (TEB), DFT-modeled XPS line shapes predict the appearance of a low BE shoulder for the organosilver TEB-Ag-TEB dimer, while the covalent TEB-TEB dimer does not exhibit this feature (\cf Figure 6b of ref. \cite{Zhang2012}). Based on the related DFT-calculation, the experimentally observed low BE shoulder is considered as a spectroscopic evidence for organometallic bonding \via Ag-bis-acetylide linkage. The argument is also in line with Kung \etal who reported a binding energy of \SI{283.7}{eV} for unsaturated methylacetylide compounds \cite{Kung2008} and studies of tessalation structures \cite{Zhang2018} and long-range ordered networks featuring Ag-bis-acetylides \cite{Zhang2019}.

\begin{figure}
	\begin{center}
		\includegraphics[width=0.9\textwidth]{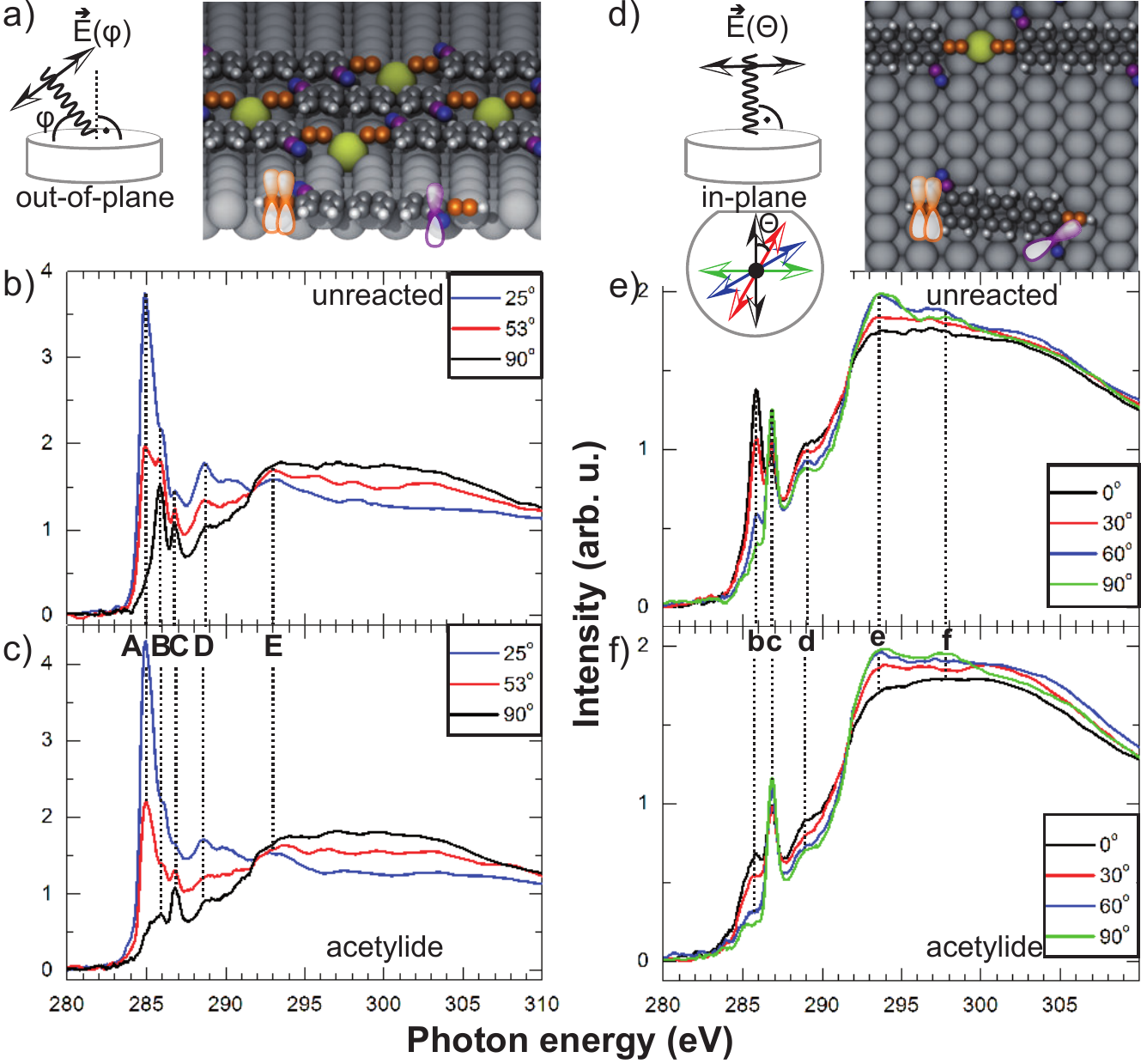}
	\end{center}
	\caption{C1s NEXAFS analysis of polar (a-c) and azimuthal (d-f) angular dependence for the pristine (b,e) and organometallic phase (c,f). (a,d) Schematic sketch depicting polar (azimuthal) NEXAFS geometry according to variations of out-of-plane (in-plane) angle $\varphi$ ($\theta$) with respect to electrical field vector $\vec{E}$. Perspective and top view of an illustrative model emphasizing out-of-plane and in-plane $\pi^*$ orbitals of C(sp) atoms within \ce{C\bond{3}N} and \ce{C\bond{3}C} groups (only shown for a monomer). Color-coding: C(sp) within alkynes (cyano groups): orange (purple); C(sp$^2$): dark grey; N: blue; H: white. (b,c) Polar NEXAFS dichroism for the pristine phase (above) and organometallic phase (below). (e,f) Azimuthal NEXAFS dichroism for the pristine phase (above) and organometallic phase (below). Dashed lines with capitalized (lowercase) letters refer to peak positions of polar (azimuthal) NEXAFS spectra.}
	\label{NEXAFS}
\end{figure}
 
Complementary to XPS, synchrotron-based NEXAFS spectroscopy is utilized to elucidate geometric, conformational, and organometallic bonding aspects within CN-func\- tionalized Ag-bis-acetylide chains. Figure \ref{NEXAFS}a depicts the geometry of the polar NEXAFS experiment, where the sample is rotated such that the incident angle $\varphi$ amounts to ($\varphi=\SI{25}{\degree}$), magical (\SI{53}{\degree}) and normal (\SI{90}{\degree}) incidence. According to the "building block" principle, \cite{Stoehr1992} each functional group within the monomeric and oliogmeric phase contributes to the recorded C1s NEXAFS profiles. C1s NEXAFS resolves the orthogonal $\pi^*$ orbitals of C(sp) atoms within terminal alkynes (orange orbitals) and cyano endgroups (purple orbitals), as indicated by the illustrative model showing pristine monomers and reacted Ag-acetylide chains aligned on Ag$(110)$ (Figure \ref{NEXAFS}a,d). They are subdivided into out-of-plane $\pi^*$ orbitals oriented perpendicular to flat-lying phenyl rings (perspective view on left side) and in-plane $\pi^*$ orbitals (top view on right side). 

The C1s NEXAFS spectra probing the directionality of the out-of-plane orbitals are shown for the organic and organometallic phase in Figure \ref{NEXAFS}b and c, respectively. The extended profiles display four sharp $\pi^*$ resonances resolved at the photon energies \SIlist{284.9;285.9;286.8;288.6}{eV}, as labeled by the capitalized letters A, B, C, D. The high photon energy window above \SI{290}{eV} is indicative for the $\sigma^*$ region with the characteristic overlapping of broadened $\sigma^*$ peaks (feature E at \SI{293.3}{eV}). A qualitative comparison of the three profiles reveals a strong dichroism for the four $\pi^*$ peaks, while the $\sigma^*$ region exhibits the opposite angle dependency. Based on comparison with the NEXAFS study by Solomon \etal providing peak positions for a sub-ML amount of benzene on Ag(110) (\{$\pi^*_{1,2}$, $\pi^*_3$, $\sigma^*_1$, $\sigma^*_2$\}=\SIlist{284.9;288.6;293.4;301}{eV}), the features A and D can be assigned to $\pi^*$ orbitals of the C(sp$^2$) atoms of the molecular backbone.
For normal incidence, feature A, which contains solely out-of-plane $\pi^*$ resonances is nearly quenched, indicating that the phenylene backbone is adsorbed coplanar with the surface. The almost constant intensity of the features B and C indicates a significant contribution from in-plane $\pi^*$ orbitals, which stem from C(sp) atoms. Since the peak position of B is consistent with previous results for a related alkyne derivative \cite{Zhang2015}, and peak position C agrees with prior studies for para-substituted dicarbonitrile-oligophenyls \cite{Klappenberger2011}, we assign them to excitations within the C(sp) atoms of the alkyne and cyano species, respectively. 

The spectra of the organometallic phase express a high degree of similar behavior regarding peaks A,C,D and E compared to the organic phase. However, taking the cyano-related resonance C as reference, a prominent change of peak B becomes obvious. The intensity is reduced, most prominently for the \SI{90}{\degree} curve, suggesting that the alkyne's in-plane $\pi^*$ orbitals got at least partially filled as a result of hybridization with orbitals of the embedded Ag centers upon organometallation.

In order to probe the azimuthal alignment of the organometallic chains on the surface and to inspect the in-plane $\pi$-orbitals of cyano and alkyne groups in more detail, further C1s NEXAFS measurements are carried out under normal incidence and varying the sample's in-plane orientation (azimuthal geometry). According to the schematic sketch in Figure \ref{NEXAFS}d, X-rays arrive under perpendicular incidence on the sample (black profiles in Figure \ref{NEXAFS}b,c). Since the $[1\overline{1}0]$ direction along the furrows of Ag(110) is known, the single crystal was initially aligned with the electrical field vector $\vec E$ (angle $\theta=\SI{0}{\degree}$). By rotating the sample in \SI{30}{\degree} steps, we probed the directionality of the in-plane orbitals, too. The relative angle $\theta$ between $\vec E$ and the $[1\overline{1}0]$ direction of the atomic lattice orientation is depicted by color-coded arrows (Figure \ref{NEXAFS}d) that refer to the C1s NEXAFS spectra of Figure \ref{NEXAFS}e,f. Consistent with the former peak assignment (B-E), the letters b, c, d, e and f represent the resonances at the photon energies \SIlist{285.9;286.9;289.0;293.9;297.5}{eV}.Taking into account the flat adsorption configuration and the nomral incidence ($\varphi=\SI{90}{\degree}$), the features b and c originate from in-plane $\pi^*$ orbitals only. The dichroism in Figures \ref{NEXAFS}e, f is related to the averaged orientations of the respective orbitals.

Since the most intense NEXAFS peak A is absent for $\varphi=\SI{90}{\degree}$ (Figure \ref{NEXAFS}b), one recognizes that feature b in Figure \ref{NEXAFS}e, f shows the highest signal for $\theta=\SI{0}{\degree}$ and decreases monotonously with increasing $\theta$ until it almost completely vanishes at $\theta=\SI{90}{\degree}$. Such behavior occurs only when the alkyne's in-plane $\pi^*$ orbitals are aligned along $\vec{v}$. This implies for both the pure and reacted phase, that monomers and organometallic wires preferentially align along $\vec{u}$, which is perpendicular to these orbitals. This conclusion is consistent with our STM results limited to microscopic sample regions. According to its linear bonding geometry, the organometallic chains represent 1D arrays of $\sigma$-like metal alkynyl complexes \cite{Long2003}.

From the nearly constant dichroism of the cyano-related signature c, we conclude that the conformation of the cyano moieties, and hence molecular alignment has not significantly changed upon the conversion from 1D ribbons to dense-packed domains. For both phases, the STM results reveal the presence of directional N$\cdots$H interactions (double hydrogen bonding) between isochiral trans isomers (\cf Figure \ref{organic}c and Figure \ref{organometallic}e). While LT-STM resolved the enantiomeric character of monomers and oligomers (D- and L-trans), NEXAFS experiments space-average over both chiral species.  Since the cyano moieties of both D-trans and L-trans isomers confine an angle of \SI{120}{\degree}, there is no field vector alignment $\theta$ that entirely quenches the contribution from one conformer species. 

Compared to the nearly unchanged angular behavior of the CN-related peak c upon sample annealing, the NEXAFS spectra in Figure~\ref{NEXAFS}f reveal a drastically attenuated and broadened peak b stemming from the alkynes' in-plane $\pi^*$ orbitals next to the embedded Ag atoms. As for the polar angular dependence in Figure \ref{NEXAFS}b and c, the azimuthal peak-ratio of \textit{b} strongly decreases (3.41$\rightarrow$2.71). We explain the drastic change of signature b by a strong hybridization between in-plane $\pi^*$ orbitals and orbitals of Ag atoms involved in organometalic C(sp)-Ag-C(sp) bonding. The hybridization-related broadening is attributed to lateral orbital overlap due to organometallic $\sigma$ bonding, while hybridization effects due to increased surface-molecule interactions within metallopolymer chains are neglected due to the nearly unchanged appearance of the CN-related NEXAFS peaks labeled c and C. From the latter, we suspect that the conformation of the cyano-phenylene moieties, and hence molecular alignment, has not significantly changed upon on-surface formation of metallopolymer wires regularly stacked within 2D domains. Furthermore, the minute concentration of covalent byproducts (explained in SI; \cf orange and purple outlines in Figure \ref{organometallic}a,f and SI Figure S2) with deviating orientations is neglected, in agreement with the cyano-related NEXAFS peaks C and c hardly showing any changes after the on-surface coupling reaction.

From the electronic configuration of Ag carrying an occupied d shell and one s electron ($[Kr]4d^{10}5s^{1}$), the linear bonding geometry is intuitively explained by $\sigma$-bonding between the $5s^1$ orbital of Ag and nearby sp orbitals of the outermost C$_{sp}$ atom of the terminal alkyne. The strong overlap of broadened $\sigma^*$ states however inhibits a deconvolution of relevant contributions in Figure \ref{NEXAFS}e,f. 

\section{Conclusions}
In summary, we report the on-surface formation and characterization of cyano-function\- alized Ag-bis-acetylide nanowires assembling into periodic 2D domains due to site-selective interchain attractions from secondary functional groups. While low-temperat-ure STM was utilized to detect adsorption aspects and isomeric states of both the CN-DETP monomer and organometallic phase at the single-molecule level, XPS unequivocally confirmed the predominance of Ag-acetylide structures on the sample. Azimuthal and polar NEXAFS measurements resolving the orbitals of primary and secondary functional groups (terminal alkynes and CN groups) not only deliver indications on oligomer conformation and alignment, but also indicate the engagement of the in-plane $\pi^*$ orbital of the alkyne's $\pi$-system in their bonding to Ag atoms.The Ag-bis-acetylide nanowires embedded within enantiomeric domains reveal an appreciable thermal stability. Accordingly, the results of our methodology provide rich impetus toward the controlled growth of functionalized and organometallic interfaces \via orthogonal interactions.

\section{Methods}
The STM-based preparations and measurements were carried out in the ultra-high vacuum environment (base pressure below \SI{1e-10}{mBar}) of a commercial Joule-Thomp-son-Scanning-Tunneling-Microscope system (\textit{SPECS GmbH}). To this end, a smooth and clean Ag(111) crystal was established by multiple sequences of Ar$^+$ sputtering (\SI{0.9}{kV}, \SI{10}{\micro\ampere} sputter current, \SI{25}{min}) and subsequent annealing at \SI{750}{K} for \SI{15}{min}. After careful degassing procedures, the molecules were deposited from an evaporation cell kept at 500~K in vacuum. The ribbon phase of pristine monomers was prepared by deposition on the clean Ag(111) sample kept at $T\le\SI{240}{K}$. While organometallic dimers and trimers formed already after annealing the pristine phase at room temperature, dense-packed domains of organometallic chains were obtained by either annealing the pristine phase above room temperature (\SIrange{350}{400}{K}), or depositing the CN-DETP molecules on the hot sample (\SI{350}{K}). STM images were acquired with a tungsten tip at a temperature of \SI{4.5}{K} and within the constant-current mode. Manipulation measurements were conducted within the open-feedback mode, whereby a sharp and stable STM tips was carefully approached to the pristine surface, followed by concerted push and pull forces on monomers and chain terminations. Synchrotron radiation experiments were carried out at the Helmholtz-Zentrum Berlin (HZB).

\section{Acknowledgement}
Funding provided by the ERC Advanced Grant MolArt (grant n\textsuperscript{o} 247299), the TUM Institute of Advanced Studies, and the German Research Foundation \via KL 2294/6-1 is gratefully acknowledged. We thank HZB for the allocation of synchrotron radiation beamtime.

\section{Supporting Information Available}
STM tip-assisted manipulation on chain termination revealing CN-related depressions and bending flexibility of a nanowire section.
Covalent byproducts and thermal stability of organometallic structures.

\newpage
\printbibliography

\end{document}


\title{Supporting information:\\Cyano-functionalized Ag-bis-acetylide wires on Ag(110)}
\author[1]{Raphael Hellwig}
\author[1]{Martin Uphoff}
\author[1]{Yiqi Zhang}
\author[2]{Ping Du}
\author[2,3]{Mario Ruben}
\author[1,4]{Florian Klappenberger}
\author[1,5]{Johannes V. Barth}
\affil[1]{Physics Department E20, Technical University of Munich, D-85748 Garching, Germany}
\affil[2]{Institut für Nanotechnologie (INT), Karlsruher Institut für Technologie (KIT), D-76344 Eggenstein-Leopoldshafen, Germany}
\affil[3]{Institut de Physique et Chimie de Mat\'{e}riaux de Strasbourg (IPCMS), CNRS-Universit\'{e} de Strasbourg, F-67034 Strasbourg, France}
\affil[4]{florian.klappenberger@tum.de}
\affil[5]{jvb@tum.de}
\date{}
\maketitle

\begin{figure}
\centering
\includegraphics[width=13cm]{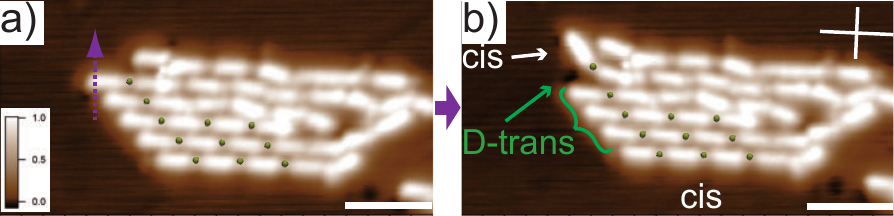}
\caption[\textbf{Figure S1.} STM tip-assisted manipulation on molecular chain.]{
STM tip-assisted manipulation on molecular chain reveals its bending flexibility.
(a) To displace the chain termination, the STM tip is guided along the depicted dashed arrow. 
Incorporated Ag atoms are emphasized by yellow spheres.
(b) STM image after manipulation displays a bent Ag-bis-acetylide motif, whereby the depression is attributed to a CN unit released from interchain bonding. 
Depression at green arrow indicates a D-trans isomers. 
From the unit cell orientation of the superlattice of embedded Ag atoms, it follows that all chain molecules within curly bracket (green) represent D-trans isomers. 
At the domain edges, the absence of depressions next to the molecules implies that they favor a cis configuration (white arrow) in order to maximize CN-mediated sidechain interactions.
Tunneling parameters $V_t$, $I_t$: (a,b) $0.01$~V, $0.1$~nA.
Scale bars: (a,b) $4$~nm.
}
\end{figure}

\section{Covalent byproducts and thermal stability of organometallic wires}

\begin{figure}
\centering 
\includegraphics[width=\textwidth]{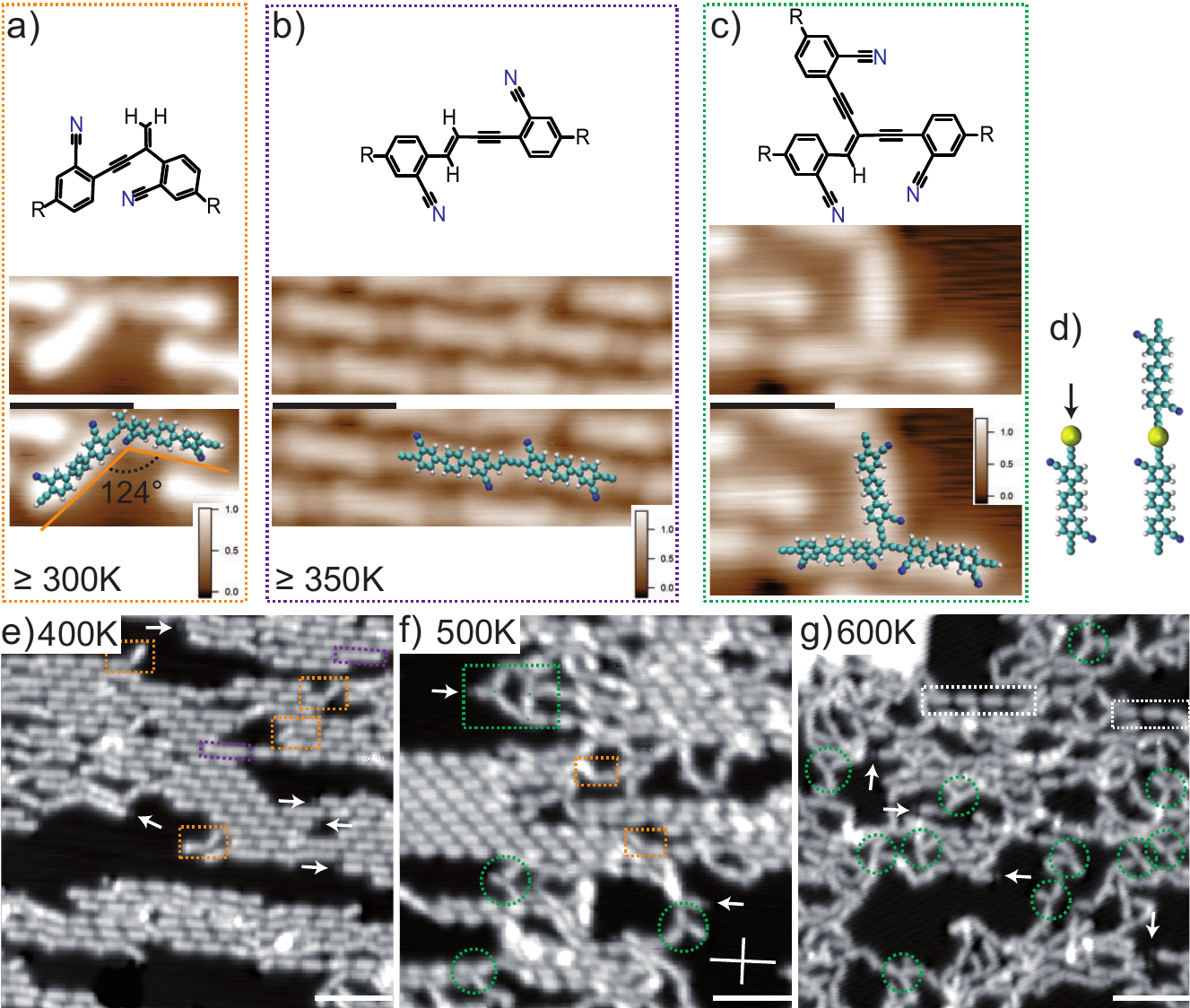}
\caption[\textbf{Figure S2.} Covalent on-surface reaction products]{\linespread{1} \normalsize
Covalent byproducts and thermal stability of Ag-acetylide motifs. 
(a-c) Proposed chemical structure (top part) according to STM images of covalent coupling motifs (middle part), along with suggested gas-phase ball-and-stick models (bottom part) superimposed to STM images.
Color-coded STM contrast helps in identifying non-interacting cyano groups.
Within chemical structure, R represents residual structure of CN-DETP monomer.
(a,b) Covalent bond formation \via vinyl-ethynyl dimerization.  
(c) Covalent trimer.
(d) Schematic representation of the mono- and bis-acetylide complex. 
High-temperature sample annealing causes increase of covalent byproducts.
(e) $400$~K: Organometallic domains with dimer products (orange and purple outline).
(f) $500$~K: Dissolvement of domains and trifurcated trimer products (green outline).
(g) $600$~K: Irregular polymeric network with trifurcation motifs and sporadic presence of bis-acetylide wires (white outline).
Tunneling parameters $V_{\textrm t}$, $I_{\textrm t}$: (a) $-0.002$~V, $0.1$~nA; (b) $0.005$~V, $0.3$~nA; (c) $0.05$~V, $0.1$~nA; (e) $0.01$~V, $0.1$~nA; (f) $0.8$~V, $0.8$~nA; (g) $0.01$~V, $0.1$~nA.
Scale bars: (a-c) $2$~nm; (e-g) $6$~nm.
}
\label{covalent}
\end{figure}

Besides organometallic chain formation, we observe bonding motifs characteristic for covalent alkyne coupling.
The frequency of the latter increases with increasing annealing temperature.
The linear C-Ag-C complex may represent an intermediate for covalent linkage \via butadiyne bridges (homocoupling), whereby there exists uncertainty whether the heat-induced split-up of Ag-acetylide chains allows the reversible uptake of another Ag atom (exchange process), bis-acetylide $\rightarrow$ mono-acetylide conversion, or exclusively leads to radical recombination \via covalent bonding.
Interestingly, after successive annealing steps within the temperature regime $\SI{350}{K}\le$T$\le\SI{600}{K}$, we clearly observe mono-acetylide complexes (C(sp)-Ag motif in Figure \ref{covalent}d), as indicated by the white arrows in Figure \ref{covalent}e-g.
This bonding scenario could either be established prior to C-Ag-C formation or/and after C-Ag-C scission processes.
In the latter case, the scission of two C(sp)-Ag-C(sp) motifs may lead to two Ag-C(sp) motifs and two C(sp) radicals immediately forming a covalent dimer. 

The reaction products not attributed to organometallic bonding are presented with a color-coded STM contrast in Figure~\ref{covalent}a-c, in order to recognize the presence/absence of depressions reflecting unbound cyano groups.
After room temperature annealing, we observe a minute concentration of an on-surface formed dimer species with angled geometry, as shown within the orange outlines in Figure~\ref{covalent}a,e,f.
Based on the good agreement with the superimposed molecular model in Figure~\ref{covalent}a, we suggest a vinyl-ethynyl coupling motif (see chemical structure for coupling motif in the top part of Figure~\ref{covalent}a).
Consistently, a further regioisomer of vinyl-ethynyl bonding is encountered, and marked by purple outlines within Figure~\ref{covalent}b,e.
From the superimposed Ag(110) registry in the manuscipt (MS) Figure~3f, it becomes evident that the Ag atoms attached to the left and right end of the covalent dimer (purple outline) are separated by nine lattice constants (36.8 \AA) along the $[001]$ direction, which is one lattice constant less compared to an organometallic dimer with terminal Ag atoms.
The high-resolution STM image of the covalent dimeric motif (formed at T~$\geq$~350~K) in Figure~\ref{covalent}b nicely agrees with the superimposed gas-phase model, \ie the displacement of the molecular axis due to the vinyl-ethynyl linkage (see chemical structure) is clearly expressed.
Straight-line motifs reminiscent of homocoupling are absent.

Threefold coupling motifs are rare for temperatures T~$\le$~400~K, but become more frequent for sample annealing at T~$\ge$~500~K (green outline in Figure~\ref{covalent}f,g).
For the species shown in Figure~\ref{covalent}c, a threefold vinyl-ethynyl-ethynyl motif is postulated according to depicted chemical structure.
While sample annealing at 400~K triggers a small amount of vinyl-ethynyl bonds (orange and purple outlines in Figure~\ref{covalent}e) within ordered domains, heating at 500~K causes gradual dissolution of dense-packed aggregates (Figure~\ref{covalent}f).
The structure conversion at domain borders is characterized by irregularly branched chains with threefold coupling nodes (green outline).
\\\newline
Subsequent to sample annealing at 600~K (Figure~\ref{covalent}g), we observe the complete conversion into a disordered polymer network.
The dominance of threefold coupling nodes and fourfold cross-coupling patterns within irregularly branched structures~\cite{Eichhorn2013} may be caused by interchain reactions between Ag-acetylide motifs.
A closer look in Figure~\ref{covalent}g reveals that mono-acetylide motifs (white arrows) and even isolated bis-acetylide wires (white outline) are preserved, thus evincing a high thermal stability of the Ag-acetylides bond with CN-DETP.
The results show that an ordered organometallic structure does not necessarily represent the intermediate toward a topologically equivalent covalent framework.
Within a related study reporting on the on-surface coupling of acetylene molecules toward Cu-bis-acetylide chains on Cu(110), high-temperature annealing also leads to the degradation of the linear structure instead of the conversion into linear covalent chains.~\cite{Sun2016a}
According to the DFT-calculated homocoupling mechanism on Ag(111) of Björk \etal predicting that the coupling step precedes the deprotonation step,~\cite{Bjoerk2014} the absence of homocoupling in our case could be related to alkyne deprotonation prior to intermolecular coupling on Ag(110).

\printbibliography